\documentclass[12pt,preprint]{aastex}

\slugcomment{version 2007.11.12}

\shorttitle{AKARI observation of SN2006jc}
\shortauthors{Sakon et al.}

\begin{document}
\title{Properties of newly formed dust by SN2006jc \\
based on near-to-mid infrared observation with {\it{AKARI}}}

\author{Sakon,I.\altaffilmark{1}, Wada,T.\altaffilmark{2}, Ohyama,Y.\altaffilmark{2}, Ishihara,D.\altaffilmark{2}, Tanab\'{e},T.\altaffilmark{3}, Kaneda,H.\altaffilmark{2}, Onaka,T.\altaffilmark{1}, \\
Tominaga,N.\altaffilmark{1}, Tanaka,M.\altaffilmark{1}, Suzuki,T.\altaffilmark{1}, Umeda,H.\altaffilmark{1}, Nomoto,K.\altaffilmark{1,4},\\
Nozawa,T.\altaffilmark{5}, Kozasa,T.\altaffilmark{5}, Minezaki,T.\altaffilmark{3}, Yoshii,Y.\altaffilmark{3},\\
Ohyabu,S.\altaffilmark{2}, Usui,F.\altaffilmark{2}, Matsuhara,H.\altaffilmark{2}, Nakagawa,T.\altaffilmark{2}, Murakami,H.\altaffilmark{2}}

\altaffiltext{1}{Department of astronomy, School of Science, University of Tokyo, 7-3-1 Hongo, Bunkyo-ku, 
Tokyo 113-0033, Japan; isakon@astron.s.u-tokyo.ac.jp.}
\altaffiltext{2}{Institute of Space and Astronautical Science, 
Japan Aerospace Exploration Agency, 3-1-1 Yoshinodai, Sagamihara, 
Kanagawa 229-8510, Japan}
\altaffiltext{3}{Institute of Astronomy, Graduate School of Science, University of Tokyo, 2-21-1 Ohsawa, Mitaka, Tokyo 181-0015, Japan}
\altaffiltext{4}{Institute of the Physics and Mathematics for the Universe, University of Tokyo, Kashiwa, Chiba 277-8568, Japan}
\altaffiltext{5}{Department of Cosmosciences, Graduate School of Science, Hokkaido University, Sapporo 060-0810, Japan}

\begin{abstract}
We present our latest results on near- to mid- infrared observation of SN2006jc  at 200 days after the discovery using the Infrared Camera (IRC) on board $AKARI$. The near-infrared (2--5$\mu$m) spectrum of SN2006jc is obtained for the first time and is found to be well interpreted in terms of the thermal emission from amorphous carbon of 800$\pm 10$K with the mass of $6.9\pm 0.5 \times 10^{-5}M_{\odot}$ that was formed in the supernova ejecta. This dust mass newly formed in the ejecta of SN 2006jc is in a range similar to those obtained for other several dust forming core collapse supernovae based on recent observations (i.e., $10^{-3}$--$10^{-5}$$M_{\odot}$). Mid-infrared photometric data with {\it{AKARI}}/IRC MIR-S/S7, S9W, and S11 bands have shown excess emission over the thermal emission by hot amorphous carbon of 800K. This mid-infrared excess emission is likely to be accounted for by the emission from warm amorphous carbon dust of 320$\pm 10$K with the mass of 2.7$^{+0.7}_{-0.5} \times 10^{-3}M_{\odot}$ rather than by the band emission of astronomical silicate and/or silica grains. This warm amorphous carbon dust is expected to have been formed in the mass loss wind associated with the Wolf-Rayet stellar activity before the SN explosion. Our result suggests that a significant amount of dust is condensed in the mass loss wind prior to the SN explosion. A possible contribution of emission bands by precursory SiO molecules in 7.5--9.5$\mu$m is also suggested. 
\end{abstract}
\pagebreak

\section{INTRODUCTION}
Study on the dust formation in the ejecta of core-collapse supernovae (SNe) is an important topic to explore the origin of dust in the early universe. Besides the theoretical studies that suggest the dust condensation in the ejecta of core-collapse SNe \citep{koz91,tod01}, several pieces of the observational evidence for the dust formation in SN ejecta have so far been reported. For example, up to 0.02$M_{\odot}$ of dust formation, which is close to the value of 0.1$M_{\odot}$ needed for core-collapse supernovae to account for the dust content of high redshift galaxies \citep{mor03}, in the ejecta of the type-II supernova 2003gd has been reported by \cite{sug06}. As for the case of the type-II supernova 1987A, however, even the highest estimate of the condensed mass in the SN ejecta reaches only up to 7.5$\times 10^{-4} M_{\odot}$ \citep{erc05}. Moreover, \cite{mei07} revised the mass of newly condensed dust in the ejecta of SN 2003gd as 4$\times 10^{-5} M_{\odot}$ based on recent observations with {\it{Spitzer}}. {\it{Spitzer}} Miltiband Imaging Photometer (MIPS) observations of the Galactic core-collapse supernova remanant Cas A show that the dust mass associated with it is much smaller ($\sim 0.003M_{\odot}$) than previously thought \citep{hin04}, while a recent {\it{Spitzer}} Infrared Spectrofraph (IRS) mapping observation \citep{rho07} reports that the total mass formed in Cas A should be at least 0.02$M_{\odot}$. There still remains a gap in the produced dust mass in core-collapse SN ejecta between those observational results and theoretical prediction of 0.1-1$M_{\odot}$ \citep{noz03}.

Supernova (SN) 2006jc is a peculiar Type Ib supernova and was discovered on 2006 October 9.75 (UT) \citep{nak06}. It is believed that the progenitor star had experienced a luminous outburst similar to those of luminous blue variables (LBVs) 2 years prior to the supernova event \citep{pas07,fol07}. Evidence of substantial interaction of the SN ejecta with a dense He-rich circumstellar medium (CSM) ejected during the LBV-like eruption was reported based on {\it{Chandra}} X-ray observations \citep{imm06}.

Recent observational studies on this supernova have shown that SN2006jc is an interesting target for the study of dust forming massive-star supernovae. The near-infrared re-brightening of SN2006jc was firstly reported by \cite{ark06} from late November through early December 2006 ($\sim 50$ days after the discovery) based on $J,H,$ and $K$ band observations. \cite{smi07} report that both the appearance of a strong continuum emission at red/near-infrared wavelengths and the fading of redshifted sides of the narrow He I emission lines occurred simultaneously between 51 and 75 days after the brightness peak. These characteristics are interpreted as the evidence for dust formation in SN2006jc, although the timescale is much shorter than the general dust formation timescale, at least a few hundred days, typical of other dust forming SNe. They found that graphite grains with $T\sim 1600$K or slightly hotter silicate grains well fit the optical(red) spectrum of day 75, pointing out that the dust is mainly carbon and not silicate because of its high temperature. 

While most of the dust forming supernovae that have been observationally reported are type-II except for the case of type-Ib SN 1990I \citep{elm04}, the type-Ib SN 2006jc gives us a unique opportunity to investigate the dust formation not only in the SN ejecta but also in rich circumstellar materials that have come from the mass loss events prior to final core-collapse. Near- to mid-infrared observations of SN2006jc in the early phase of dust condensation are essential to examine the composition and properties of the newly formed dust in the SN ejecta and the pre-existing dust in the circumstellar medium separately. In this paper we present the new near-infrared spectrum and the mid-infrared photometric data of SN2006jc taken with the Infrared Camera(IRC) \citep{ona07} on board {\it{AKARI}} \citep{mur07} on 29 April 2006 (200 days after the discovery). We derive the properties of the carriers of re-brightened near-infrared emission and the mid-infrared emission to understand the dust formation in SN2006jc.

\pagebreak

\section{Observation and Data Reduction}
Two pointed observations of SN2006jc were performed at 00:36:22(UT) and 02:15:47(UT) on 29 April 2007 as part of the director's time of {\it {\it{AKARI}}}. The former observation (ID:5124071) was performed with the AOT04 mode, in which the mid-infrared spectroscopic data were taken with two grisms, SG1 (5.4--8.4$\mu$m) and SG2 (7.5-12.9$\mu$m), and the mid-infrared imaging data were taken with the broad band filter, S9W(6.7$\mu$m--11.6$\mu$m) of the MIR-S channel and the near-infrared spectroscopic data were taken with the grism, NG (2.5--5.0$\mu$m), and the near-infrared imaging data were taken with the broad band filter, N3(2.7--3.8$\mu$m) of the NIR channel. The latter observation (ID:512472) was performed with AOT02b mode, in which the mid-infrared imaging data were taken with two medium band filters, S7 (5.9--8.4$\mu$m) and S11 (8.5--11.3$\mu$m) of the MIR-S channel and simultaneously the near-infrared spectroscopic data was taken with the prism, NP (1.8--5.2$\mu$m), and the near-infrared imaging data were taken with the broad band filter, N3(2.7--3.8$\mu$m) of the NIR channel. The target was very faint and the spectrum was seriously affected by the blending with that of the host galaxy UGC4904, which hinders us to obtain useful spectral information with NG, SG1 and SG2. In the following analysis we use the spectrum taken with NP as well as the imaging data with N3, S7, S9W, and S11. The total exposure time was 206s, 206s, 224s, 56s, and 224s for NP, N3, S7, S9W and S11, respectively. 

Each of imaging data reduction procedures of subtraction of the detector dark current, corrections for the effect of the hit by high-energy ionizing particles and the scattered light \citep[][]{sak07}, the flat fielding, the distortion correction, and the shift and co-addition of the exposure frames for N3, S7, S9W, and S11 bands follows those in the {\it IRC Imaging Pipeline Version 20070912} and the unit conversion factors from Analogue-to-Digital Converter Unit (ADU) per unit time to Jy are taken from {\it {\it{AKARI}} IRC Data Users Manual ver.1.3}. The pixel scale of the NIR channel is 1."46 and that of the MIR-S channel is 2."34. The FWHM of the image size is 4."0, 5."1, 5."5, and 4."8 and the peak central pixel flux is 5.9\%, 12.0\%, 11.5\%, and 12.1\% for NIR/N3, MIR-S/S7, S9W, and S11, respectively \citep{ona07}. Each of the spectroscopic data reduction procedures of subtraction of the detector dark current, correction for the high-energy ionizing particles effects, flat fielding, and the shift and co-addition of the exposure frames for NP data follows those in the {\it IRC Spectroscopy Toolkit Version 20070913} \citep{ohy07}. 

\begin{figure}[htbp!]
\hspace{1cm}
\includegraphics[width=0.9\linewidth]{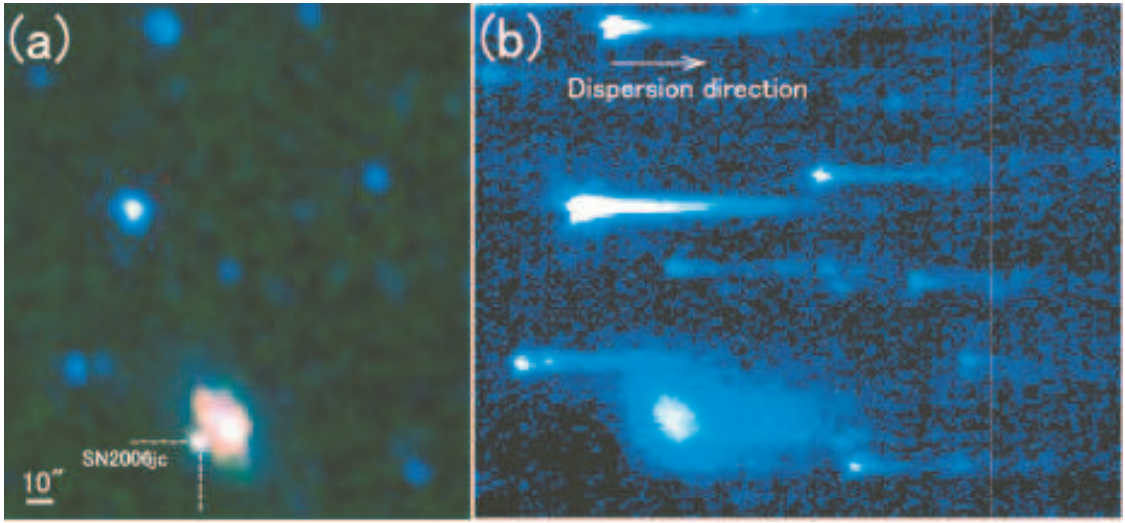}
\caption{(a) RGB false color image of SN2006jc obtained with the {\it{AKARI}}/IRC NIR/N3 (blue), MIR-S/S7 (green), and S11 (red) bands. (b) Slitless spectroscopic image with NIR/NP. The dispersion direction to longer wavelengths is shown with the white arrow. \label{f1}}
\end{figure}

 Fig.~\ref{f1}a shows the RGB false color image of SN2006jc produced with {\it{AKARI}}/IRC NIR/N3(blue), MIR-S/S7(green), and S11(red) bands, where the data taken with N3 and S11 bands are degraded into the gaussian beam with the FWHM of 5."1 so that they match with the image size of the S7 band data. SN2006jc is located close ($\sim$10") to the nucleus of the host galaxy UGC4904 and the diffuse background or foreground component of UGC4904 must be carefully removed to obtain the flux solely from the SN2006jc. The photometric decomposition technique employed is described in \S 2.1. The near infrared spectral image of SN2006jc is obtained by the slitless spectroscopy with {\it{AKARI}}/IRC NP (see Fig.1b) and it shows that the spectrum of SN2006jc is seriously blended with that of UGC4904 and a careful subtraction of the UGC4904 component is needed to derive the pure spectrum of SN2006jc. The spectroscopic decomposition technique employed is described also in \S 2.2.\\

\subsection{photometric decomposition technique}
To subtract the host galaxy component, we first derive the most appropriate point spread function (PSF) in the image and assume the galaxy component as a convolution of the source intensity distribution with PSF. This procedure is carried out in the vertical and horizontal direcion independently to check the reliability and consistensy.

\begin{figure}[htbp!]
\hspace{1cm}
\includegraphics[width=0.9\linewidth]{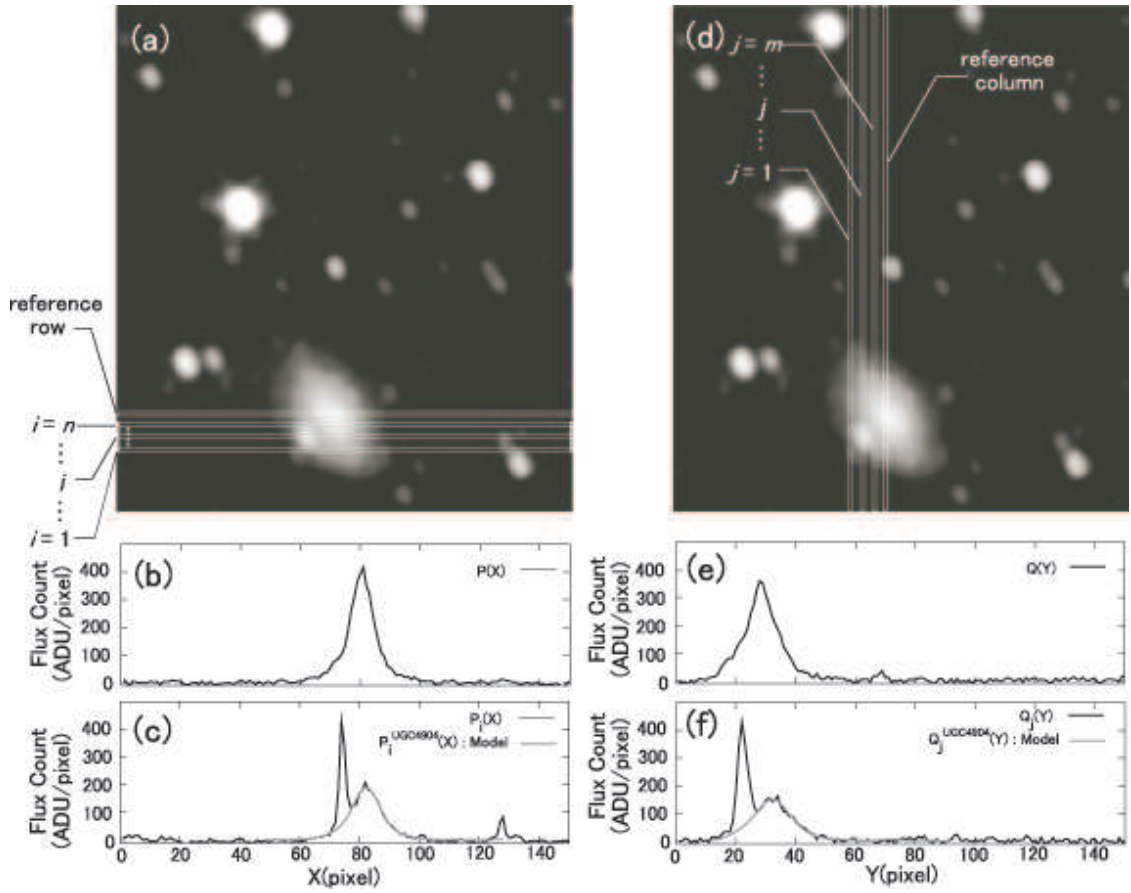}
\caption{The photometric decomposition techniques adopted for observation of SN2006jc with {\it{AKARI}}/IRC NIR/N3 band. \label{fa1}}
\end{figure}

We obtain the intensity profile $P_i(X)$ ($i=1,2,...n$) in units of ADU pixel$^{-1}$ along the $i$-th pixel row (see Fig.~\ref{fa1}c), where $n$ is defined by the number of the pixel rows that cover the SN2006jc (see Fig.~\ref{fa1}a). We also define the reference pixel row which goes across only the host galaxy UGC4904 and obtain the intensity profile $P(X)$ in units of ADU pixel$^{-1}$ along the reference pixel row (see Fig.~\ref{fa1}b). The position of the reference pixel row is selected so that $P(X)$ has as sharp a profile as possible and, therefore, we use $P(X)$ as the PSF. Then, we assume the intensity profile of the UGC4904 component $P^{UGC4904}_i(X)$ in $P_i(X)$ as a convolution with the PSF written as
\begin{eqnarray}
P_i^{UGC4904}(X)=\sum^{\infty}_{k=-\infty}\eta_i(k) P(X-k),
\end{eqnarray}
where the distribution function $\eta_i(k)$ is assumed to have the gaussian form given by
\begin{eqnarray}
\eta_i(k)=h_i \exp \left\{ -\left(\frac{k-\delta_i}{\sigma_i}\right) \right\}.
\end{eqnarray}
The best fit parameters of $h_i$, $\delta_i$, and $\sigma_i$ ($i=1,2,...,n$) are obtained so that the sum of $\vert P_i(X)-P_i^{UGC4904}(X) \vert^2$ for $50 \le X \le 69$ and $79 \le X \le 90$ becomes minimum, where the data of $P_i(X)$ in $70 \le X \le 78$ have contribution from the signal of SN2006jc and excluded. The intensity profile of the SN2006jc component $P^{SN2006jc}_i(X)$ is, then, calculated as
\begin{eqnarray}
P^{SN2006jc}_i(X) = P_i(X) - P_i^{UGC4904}(X).
\end{eqnarray}
Finally, the total N3 band flux estimated from the analysis on the intensity profiles along the $X$ direction, $f_{\nu}^{X}(N3)$, is derived as
\begin{eqnarray}
f_{\nu}^{X}(N3) = \alpha \sum^{n}_{i=1} \int P^{SN2006jc}_i(X) dX
\end{eqnarray}
where $\alpha$ is the unit conversion factor from ADU to Jansky.\\

In order to check the validity of the estimated flux density of SN2006jc using the intensity profiles along the $X$ direction, the same decomposition technique is applied along the $Y$ direction. We define $m$ of target pixel columns so that they cover over the whole SN2006jc (see Fig.~\ref{fa1}d) and obtain the intensity profile $Q_j(Y)$ ($j=1,2,...,m$) in units of ADU pixel$^{-1}$ along the $j$-th pixel column (see Fig.~\ref{fa1}f). We assume the intensity profile of the UGC4904 component $Q^{UGC4904}_j(Y)$ in $Q_j(Y)$ as
\begin{eqnarray}
Q_j^{UGC4904}(Y)=\sum^{\infty}_{k=-\infty}\eta_j(k) Q(Y-k),
\end{eqnarray}
where $Q(Y)$ is the PSF in the $y$ direction represented by the intensity profile along the reference pixel column which is chosen in the same way as in the $x$ direction (see Fig.~\ref{fa1}e) and the distribution function $\eta_j(k)$ is assumed to have the gaussian profile. It is obtained so that the sum of $\vert Q_j(Y)-Q_j^{UGC4904}(Y) \vert^2$ for $1 \le Y \le 18$ and $28 \le Y \le 40$ become minimum, where the data of $Q_j(Y)$ in $19 \le Y \le 27$ are contributed by the signal from SN2006jc. Then the total N3 band flux estimated from the analysis on the intensity profiles along the $Y$ direction $f_{\nu}^{Y}(N3)$ is derived as
\begin{eqnarray}
f_{\nu}^{Y}(N3) = \alpha \sum^{m}_{j=1} \int Q_j(Y) - Q^{UGC4904}_j(Y) dY.
\end{eqnarray}

The obtained N3 band fluxes $f_{\nu}^{X}(N3)$ and $f_{\nu}^{Y}(N3)$ are listed in Table.~\ref{a1}. These two values are in good agreement within uncertainties. We adopted the weighted average of the two values taking account of the error as the final N3 band flux of SN2006jc.

We obtain the flux densities of SN2006jc with MIR-S/S7, S9W, and S11 bands using the same decomposition technique as that adopted for NIR/N3 band, and the results are summarized in Table.~\ref{a1}. The flux densities estimated from the intensity profiles along the $X$ direction and $Y$ direction are in good agreement within uncertainties for any of MIR-S/S7, S9W, and S11 bands.

\begin{table}[htbp!]
\begin{center}
\caption{Decomposed flux density of SN2006jc on 29 April 2007 (200 days after the discovery). \label{a1}}
\begin{tabular}{ccc}
\hline \hline
band & $f_{\nu}^{X}$ & $f_{\nu}^{Y}$ \\ 
     &   (mJy)   & (mJy) \\   \hline
 NIR/N3 & 0.386$\pm 0.046$ & 0.376$\pm 0.042$ \\
 MIR-S/S7 & 0.586$\pm 0.063$ & 0.548$\pm 0.070$ \\
 MIR-S/S9W & 0.694$\pm 0.098$ & 0.735$\pm 0.105$ \\
 MIR-S/S11 & 0.495$\pm 0.061$ & 0.501$\pm 0.098$ \\ \hline
\end{tabular} 
\end{center}
\end{table}

\pagebreak

\subsection{spectral decomposition technique}
The near infrared spectrum of SN2006jc is obtained by the slit-less spectroscopy with NIR/NP, in which the spectrum is dispersed in the $X$ direction and a spectrum of a certain source is contaminated by the spectra of other sources aligning in the dispersion direction. Therefore, the spectrum of SN2006jc suffers severe blending with the light from UGC4904. We have developed a spectral decomposition technique for the spectral image of NIR/NP. We use the imaging data taken with NIR/N3 to investigate the structure of UGC4904 and to reproduce the spectral components of UGC4904 in the spectral image of NIR/NP, and have obtained a pure spectrum of SN2006jc. \\

\begin{figure}[htbp!]
\hspace{1cm}
\includegraphics[width=0.9\linewidth]{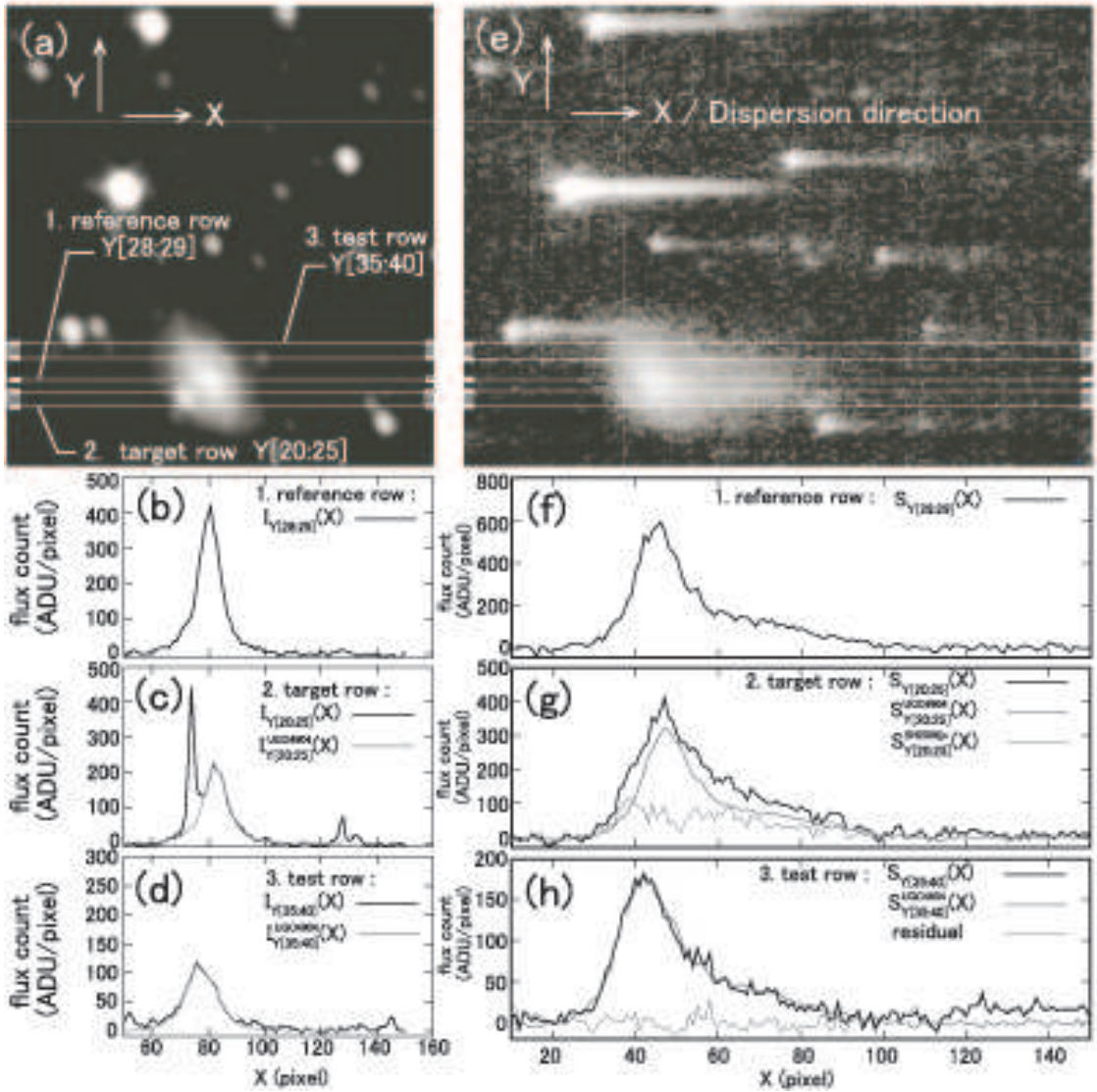}
\caption{The spectroscopic decomposition techniques adopted for observation of SN2006jc with {\it{AKARI}}/IRC NIR/NP band. \label{fa2}}
\end{figure}

For the first step, the position in the $Y$ direction of the NIR/N3 data is adjusted to that of the NIR/NP data with an accuracy of one-tenth of the pixel.

We define the reference pixel row in the imaging data taken with NIR/N3 covering from Y=28 to Y=29 so that it goes across only UGC4904 and does not include SN2006jc (see Fig.~\ref{fa2}a). The intensity profile along the reference pixel row, $I_{{\rm{Y[28:29]}}}(X)$, is obtained by averaging the data from Y=28 to Y=29 (see Fig.~\ref{fa2}b). Then, we define the target pixel row covering from Y=20 to Y=25 so that it goes across the major part of SN2006jc (see Fig.~\ref{fa2}a) and the intensity profile along the target pixel row, $I_{{\rm{Y[20:25]}}}(X)$, is obtained by averaging the data from Y=20 to Y=25 (see Fig.~\ref{fa2}c). 

\begin{figure}[htbp!]
\hspace{1cm}
\includegraphics[width=0.9\linewidth]{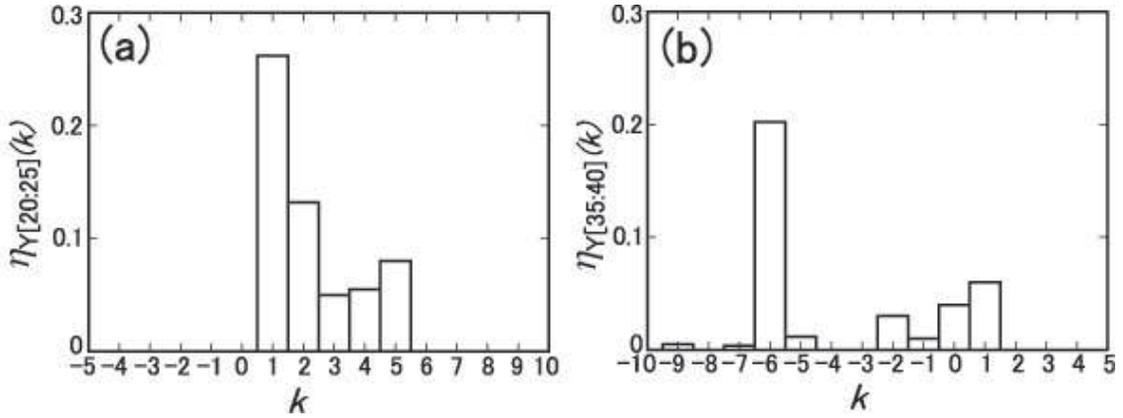}
\caption{The obtained distribution functions of (a) $\eta_{{\rm{Y[20:25]}}}(k)$ for target pixel row and (b)$\eta_{{\rm{Y[35:40]}}}(k)$ for test pixel row. \label{a3}}
\end{figure}

We model the intensity profile of the UGC4904 component, $I^{UGC4904}_{{\rm{Y[20:25]}}}(X)$, contained in $I_{{\rm{Y[20:25]}}}(X)$ with 
\begin{eqnarray}
I_{{\rm{Y[20:25]}}}^{UGC4904}(X)=\sum^{\infty}_{k=-\infty}\eta_{{\rm{Y[20:25]}}}(k) I_{{\rm{Y[28:29]}}}(X-k),
\end{eqnarray}
where the free distribution function, $\eta_{{\rm{Y[20:25]}}}(k)$, is determined so that the sum of $\vert I_{{\rm{Y[20:25]}}}(X) - I_{{\rm{Y[20:25]}}}^{UGC4904}(X) \vert^2$ for $50 \le X \le 69$ and $79 \le X \le 100$ becomes minimum, where the data of $P_i(X)$ in $70 \le X \le 78$ have contribution from the signal of SN2006jc and excluded (see Fig.~\ref{fa2}c). The obtained result of $\eta_{{\rm{Y[20:25]}}}(k)$ is shown in Fig.~\ref{a3}a.

Then we define the reference pixel row and the target pixel row in the spectral image of NIR/NP so that each has the same $Y$ range as that defined in the image of NIR/N3. The spectral intensity profiles along the reference pixel row, $S_{{\rm{Y[28:29]}}}(X)$ (see Fig.~\ref{fa2}f) and along the target pixel row, $S_{{\rm{Y[20:25]}}}(X)$ (see Fig.~\ref{fa2}g) are obtained.

Using the spectral intensity profile along the reference pixel row, $S_{{\rm{Y[28:29]}}}(X)$, and the obtained distribution function $\eta_{{\rm{Y[20:25]}}}(k)$, the spectral intensity profile of the UGC4904 component, $S^{UGC4904}_{{\rm{Y[20:25]}}}(X)$, contained in $S_{{\rm{Y[20:25]}}}(X)$ can be modeled with
\begin{eqnarray}
S_{{\rm{Y[20:25]}}}^{UGC4904}(X)=\sum^{\infty}_{k=-\infty}\eta_{{\rm{Y[20:25]}}}(k) S_{{\rm{Y[28:29]}}}(X-k),
\end{eqnarray}
which is shown with the gray line in Fig.~\ref{fa2}g. The spectral intensity profile of SN2006jc, $S_{{\rm{Y[20:25]}}}^{SN2006jc}(X)$, contained in $S_{{\rm{Y[20:25]}}}(X)$ is given by
\begin{eqnarray}
S_{{\rm{Y[20:25]}}}^{SN2006jc}(X)=S_{{\rm{Y[20:25]}}}(X)-S_{{\rm{Y[20:25]}}}^{UGC4904}(X)
\end{eqnarray}
and is shown with the thin black line in Fig.~\ref{fa2}g. 

In order to evaluate the accuracy in the estimate of the spectral intensity profile of the UGC4904 component by this spectral decomposition technique, 
we define the test pixel row both in the image of NIR/N3 and NP covering from Y=35 to Y=40 so that it goes across the host galaxy UGC4904 (see Fig.~\ref{fa2}a and e) and applied the same technique (see Figs.~\ref{fa2}d and h).

We model the intensity profile of the UGC4904 component, $I^{UGC4904}_{{\rm{Y[35:40]}}}(X)$, contained in $I_{{\rm{Y[35:40]}}}(X)$ with 
\begin{eqnarray}
I_{{\rm{Y[35:40]}}}^{UGC4904}(X)=\sum^{\infty}_{k=-\infty}\eta_{{\rm{Y[35:40]}}}(k) I_{{\rm{Y[28:29]}}}(X-k),
\end{eqnarray}
where the free distribution function $\eta_{{\rm{Y[35:40]}}}(k)$ is derived in the same way as $\eta_{{\rm{Y[20:25]}}}(k)$ (see Fig.~\ref{fa2}d). The obtained result of $\eta_{{\rm{Y[35:40]}}}(k)$ is shown in Fig.~\ref{a3}b.

The spectral intensity profile of the UGC4904 component, $S^{UGC4904}_{{\rm{Y[35:40]}}}(X)$, is given by
\begin{eqnarray}
S_{{\rm{Y[35:40]}}}^{UGC4904}(X)=\sum^{\infty}_{k=-\infty}\eta_{{\rm{Y[35:40]}}}(k) S_{{\rm{Y[28:29]}}}(X-k).
\end{eqnarray}

The obtained results of $S^{UGC4904}_{{\rm{Y[35:40]}}}(X)$ and the residual profile given by $S_{{\rm{Y[35:40]}}}(X)-S_{{\rm{Y[35:40]}}}^{UGC4904}(X)$ are shown with the gray line and the thin line in Fig.~\ref{fa2}h, respectively. The standard deviation of the residual profile from zero is 8.1(ADU/pixel), which confirms that the spectral intensity profile of the UGC4904 component is properly estimated by our spectral decomposition technique. The value of 8.1(ADU/pixel) is used to estimate the systematic error in the spectral decomposition procedure.

We have made the wavelength calibration and the division by the system spectral response of NIR/NP for the obtained spectral intensity profile $S_{{\rm{Y[20:25]}}}^{SN2006jc}(X)$ following the procedure in the IRC Spectroscopic toolkit \citep{ohy07}. The absolute flux calibration for the NP spectrum of SN2006jc has been made by using the ratio of the total flux count of SN2006jc in $I_{{\rm{Y[20:25]}}}(X)$ to that obtained by the photometric decomposition technique in \S 2.1. 
\pagebreak

\section{RESULTS}
\subsection{The near- to mid-infrared characteristics of SN2006jc}
 The results of the photometry of SN2006jc with {\it{AKARI}}/IRC N3, S7, S9W, and S11 bands are listed in Table.~\ref{t1}. Near-infrared photometry of SN2006jc with H and K bands was performed simultaneously on 28 and 29 April 2007 \citep{min07} using the multicolor imaging photometer (MIP) mounted on the MAGNUM 2m telescope at Haleakala Observatories in Hawaii \citep{kob98a,kob98b}, and the results are summarized also in Table.~\ref{t1}.

\begin{table}[htbp!]
\begin{center}
\caption{Results of the photometry of SN2006jc on 29 April 2007 (200 days after the discovery). \label{t1}}
\begin{tabular}{cccc}
\hline \hline
Instrument & band & $\lambda_{band}$ ($\mu$m) & flux density (mJy) \\ \hline
MAGNUM/MIP & $H$ & 1.6 & 0.05$\pm 0.02$ \\
MAGNUM/MIP & $K$ & 2.2 & 0.14$\pm 0.02$ \\ 
AKARI/IRC & N3 & 3.2 & 0.38$\pm 0.04$ \\
AKARI/IRC & S7 & 7.0 & 0.57$\pm 0.07$ \\
AKARI/IRC & S9W & 9.0 & 0.71$\pm 0.10$ \\
AKARI/IRC & S11 & 11.0 & 0.50$\pm 0.10$ \\ \hline
\end{tabular} 
\end{center}
\end{table}

\begin{figure}[htbp!]
\hspace{1cm}
\includegraphics[width=0.9\linewidth]{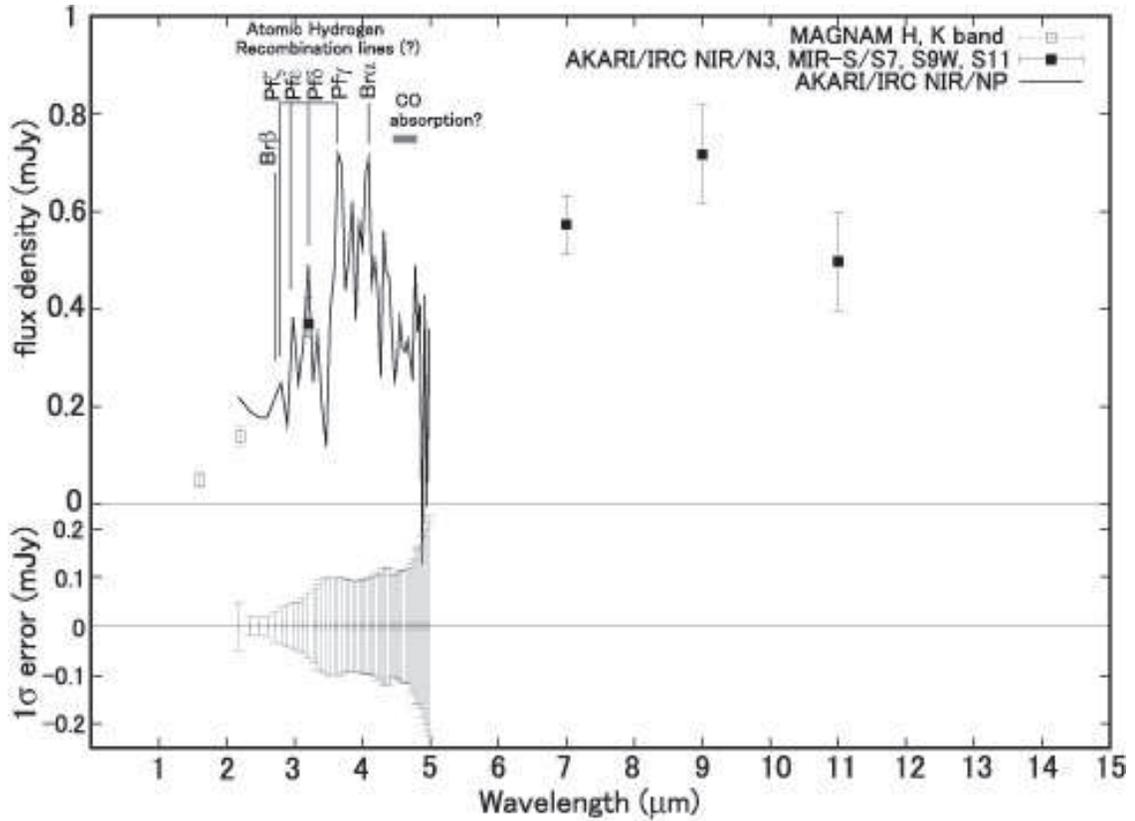}
\caption{The near-infrared spectrum of SN2006jc on day 200 after the discovery with {\it{AKARI}}/IRC NIR/NP. Photometric data taken with MAGNAM/MIP H, K bands (open square) and with {\it{AKARI}}/IRC NIR/N3, MIR-S/S7, S9W, S11 bands (solid square) are plotted together. The 1-$\sigma$ error level in the NP spectrum are also shown.  \label{f2}}
\end{figure}

The obtained near-infrared spectrum of SN2006jc on day 200 together with the photometric results are plotted in Fig.~\ref{f2}. The near-infrared spectrum of SN2006jc is characterized by a continuum emission peaking around at $\sim$4$\mu$m, which can be attributed to the thermal emission from dust grains. Possible atomic hydrogen recombination lines of Pf-$\zeta$ at 2.87$\mu$m, Pf-$\epsilon$ at 3.04$\mu$m, Pf-$\delta$ at 3.30$\mu$m, Pf-$\gamma$ at 3.74$\mu$m, Br-$\alpha$ at 4.07$\mu$m and Br-$\beta$ at 2.63$\mu$m are recognized though they are significant only within 2--3$\sigma$ (see Fig.~\ref{f2}). We also note a small dent around at 4.6$\mu$m possibly attributed to the CO absorption band. 

\subsection{Emission from the Newly-Formed Dust in the Near-Infrared}
Assuming that spherical dust grains of a certain kind $X$ with a uniform particle radius $a_{X}$ and a total mass of $M_{X}$ are located at a distance of $R$ from the observer and that they emit optically thin thermal radiation of the temperature of $T_{X}$(K), the observed flux density profile is given by
\begin{eqnarray}
f^{X}_{\nu}(\lambda)= M_X \left( \frac{4}{3}\pi \rho_{X}a_{X}^{3}\right)^{-1} \pi B_{\nu}(\lambda,T_{X}) Q^{abs}_{X}(\lambda) \left( \frac{a_{X}}{R} \right)^{2}, \label{EQ1}
\end{eqnarray}
where $Q^{abs}_{X}(\lambda)$ is the absorption efficiency and $\rho_{X}$ is the density of a dust particle of composition $X$. The near-infrared spectral energy distribution (SED) of SN2006jc on day 200 of $H$ and $K$ band photometric data and the 2--5$\mu$m spectroscopic data is fitted with $f^{X}_{\nu}(\lambda)$ for each case of the amorphous carbon ($X=a.car.$) and the astronomical silicate ($X=a.sil.$) as a composition of dust. We assume the distance of $R=25.8$Mpc \citep{pas07} for SN2006jc.

\begin{figure}[htbp!]
\hspace{1.5cm}
\includegraphics[width=0.75\linewidth]{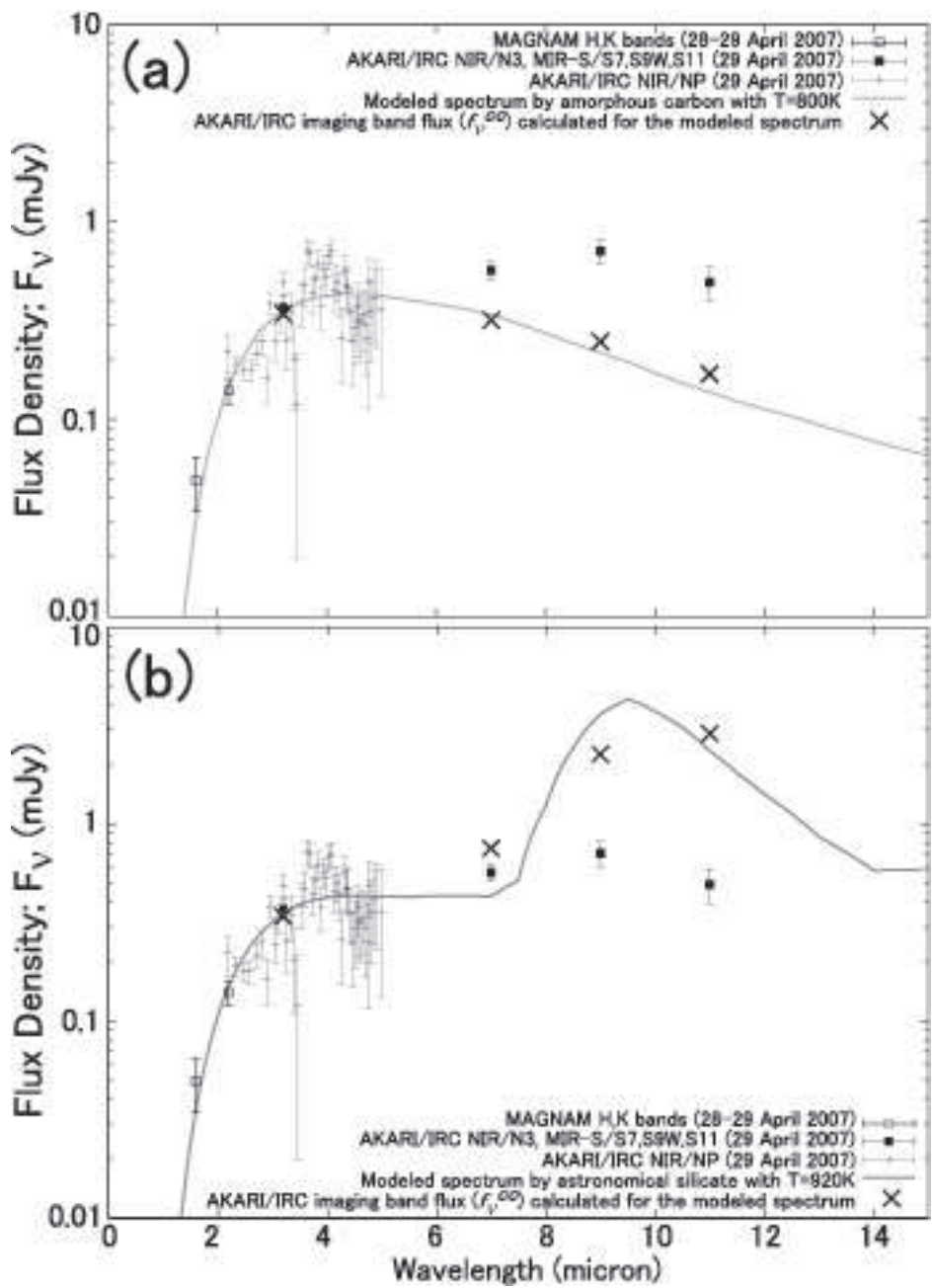}
\caption{Near-to mid-infrared spectral energy distribution (SED) of SN2006jc on day 200 constructed by the $H$ and $K$ band photometric data($open$ $square$), the 2--5$\mu$m spectroscopic data($plus$), and the mid-infrared photometric data($solid$ $square$). The results of the best fit spectrum modeled with Eq.~(\ref{EQ1}) to reproduce the near-infrared data are shown for the case of (a) amorphous carbon ($X=a.car.$) and (b) astronomical silicate ($X=a.sil.$). The flux density at {\it{AKARI}}/IRC NIR/N3, MIR-S/S7, S9W, and S11 bands simulated for the model spectrum taking account of the color correction is shown with the thick crosses.  \label{f3}}
\end{figure}

As for the amourphous carbon case, the absorption efficiency, $Q^{abs}_{a.car.}(\lambda)$, for a spherical amorphous carbon grain with a radius of $a_{a.car.}=0.01\mu$m is calculated from the optical constants of \cite{edo83}. Assuming $\rho_{a.car.}=2.26$g cm$^{-3}$, the fit is carried out with the equilibrium temperature $T_{a.car.}$ and the total mass $M_{a.car.}$ being free parameters. The result of the fit is shown in Fig.~\ref{f3}a and the best fit parameters of $T_{a.car.}=800\pm 10$ K and $M_{a.car.}=6.9\pm 0.5 \times 10^{-5}M_{\odot}$ are obtained. 

The flux density at each {\it{AKARI}}/IRC imaging band is calculated for the model spectrum taking account of the color correction, which is given by
\begin{eqnarray}
f^{a.car.}_{\nu}(band)=\frac{\displaystyle \int^{\infty}_{0} R_{band}(\nu) f^{a.car.}_{\nu}(\nu) d\nu }{\displaystyle \int^{\infty}_{0} \left( \frac{\nu_{band}}{\nu} \right) R_{band}(\nu)d\nu}, \label{EQ2}
\end{eqnarray}
where $R_{band}(\nu)$ is the relative system spectral response of each {\it{AKARI}}/IRC band and $\nu_{band}=c/\lambda_{band}$ corresponds to the reference frequency of each {\it{AKARI}}/IRC band defined with $\lambda_{N3}$=3.2$\mu$m, $\lambda_{S7}$=7.0$\mu$m, $\lambda_{S9W}$=9.0$\mu$m, $\lambda_{S11}$=11.0$\mu$m \citep{ona07}. The model values that were converted into the AKARI/IRC calibration system of $f^{a.car.}_{\nu}(N3)$, $f^{a.car.}_{\nu}(S7)$, $f^{a.car.}_{\nu}(S9W)$, and $f^{a.car.}_{\nu}(S11)$ are plotted by crosses in Fig.~\ref{f3}a. The observed data at S7, S9W, S11 bands have excess emission over $f^{a.car.}_{\nu}(S7)$, $f^{a.car.}_{\nu}(S9W)$, and $f^{a.car.}_{\nu}(S11)$, respectively. The interpretation for the excess component is discussed in \S 4.

As for the silicate case, the absorption efficiency, $Q^{abs}_{a.sil.}(\lambda)$, for a spherical astronomical silicate grain with a radius of $a_{a.car.}=0.01\mu$m is taken from the values in \cite{dra85}, and $\rho_{a.sil.}=3.3$g cm$^{-3}$ is assumed. The fit was carried out in the same way as for the amorphous carbon grains. The result of the fit is shown in Fig.~\ref{f3}b and the best fit parameters of $T_{a.sil.}=920\pm 10$(K) and $M_{a.sil.}=4.2 \pm 0.3 \times 10^{-4}M_{\odot}$ are obtained. The flux density at N3, S7, S9W, and S11 bands is calculated for the best fit model spectrum taking account of the color correction and is shown in Fig.~\ref{f3}b. The model values of $f^{a.sil.}_{\nu}(S7)$, $f^{a.sil.}_{\nu}(S9W)$, and $f^{a.sil.}_{\nu}(S11)$ largely exceed the observed flux density at S7, S9W, and S11 bands. The disagreement in the mid-infrared flux shows that astronomical silicate is not a likely carrier of the continuum in the 2--5$\mu$m region on day 200. 

We conclude that amorphous carbon is a likely carrier of the near-infrared continuum on day 200. The spectrum is well accounted for by the amorphous carbon dust with the temperature of $800 \pm 10$ K and with the total mass of $M_{a.car.}=6.9 \pm 0.5 \times 10^{-5}M_{\odot}$. This result is consistent with the suggestion made by \cite{smi07} that the re-brightened near-infrared emission on 79days is likely to be carried by carbonaceous dust with an equilibrium temperature of $1600$K, but not by silicate dust. The equilibrium temperature of 800K at on day 200 derived in our analysis is consistent with the scenario that this component is newly formed dust in the free-expanding ejecta of the supernova \citep{noz07}.

\pagebreak

\section{Identification of the Mid-Infrared Excess Component Carriers}
The mid-infrared excess component is seen over the model spectrum of the amorphous carbon of 800K at the S7, S9W and S11 bands. One possible candidate for the excess component is the dust with an emission band in the mid-infrared region carried by silicate and/or amorphous silica dust, which is discussed in \S 4.1.  Another possible candidate for the excess component is the thermal emission from the dust with temperature lower than 800K. The properties of the lower temperature dust are discussed in \S 4.2. In \S 5 we discuss which one of these two candidates is more likely. \\

 In the following analysis, we model the near- to mid-infrared spectrum as a combination of $N$ kinds of dust components ($X_i$; $i=1,...,N$) including the amorphous carbon discussed in \S 3.2 and then the model spectrum is calculated as
\begin{eqnarray}
f^{model}_{\nu}(\lambda)=\sum^{N}_{i=1} M_{X_i} \left( \frac{4}{3}\pi \rho_{X_i}a_{X_i}^{3}\right)^{-1} \pi B_{\nu}(\lambda,T_{X_i}) Q^{abs}_{X_i}(\lambda) \left( \frac{a_{X_i}}{R} \right)^{2}, \label{EQ3}
\end{eqnarray}
where the fit parameters $M_{X_i}$ and $T_{X_i}$ are the mass and the temperature of the dust component $X_i$ and $Q^{abs}_{X_i}(\lambda)$, $a_{X_i}$, and $\rho_{X_i}$ are the absorption efficiency, grain radius, and the density of dust $X_i$, respectively.

\subsection{Silicate and/or Amorphous SiO$_{2}$ Model}
\begin{figure}[htbp!]
\includegraphics[width=0.9\linewidth]{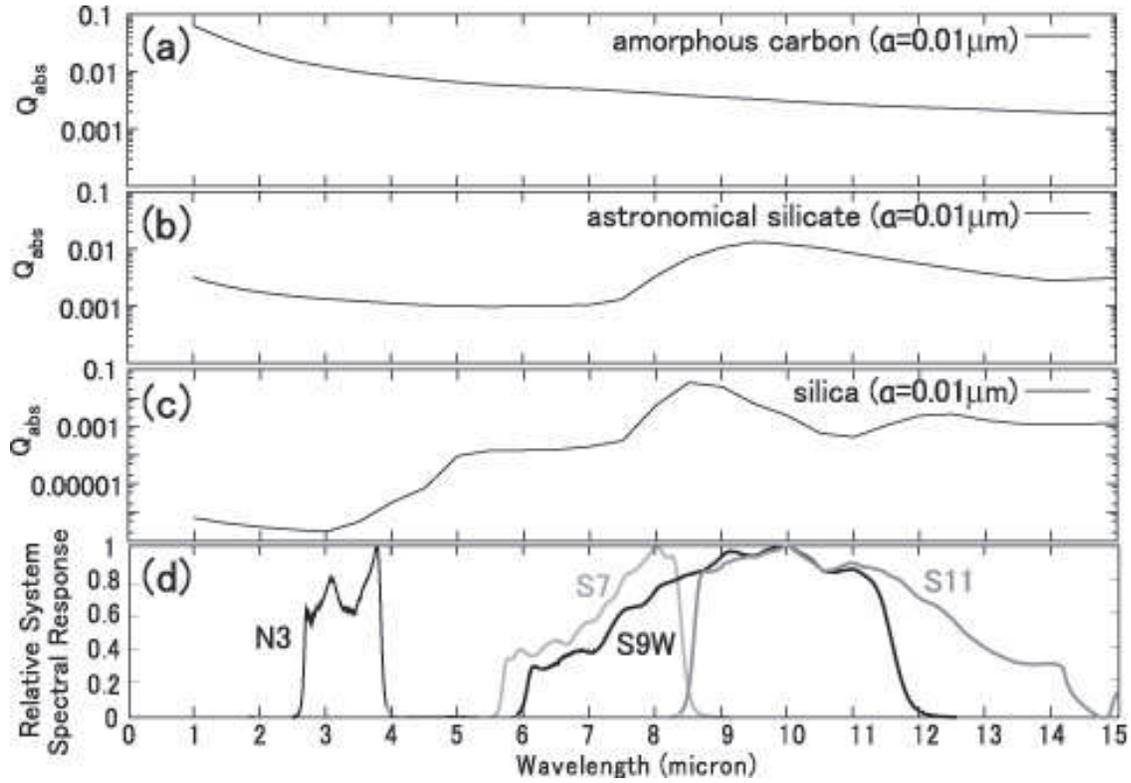}
\caption{Absorption efficiency profile $Q^{abs}_{X}(\lambda)$ of (a) amorphous carbon ($X=a.car.$), (b)astronomical silicate ($X=a.sil.$), and (c) amorphous SiO$_2$ ($X=silica$). (d) The system spectral response curve of {\it{AKARI}}/IRC NIR/N3, MIR-S/S7, S9W, and S11 bands. \label{f4}}
\end{figure}

One possible candidate for the excess emission in S7, S9W, and S11 bands is the dust band emission of the silicate and/or amorphous silica dust. In Fig.~\ref{f4} are shown the absorption efficiency profiles $Q^{abs}_{X}(\lambda)$ of amorphous carbon ($X=a.car.$) from \cite{edo83}, astronomical silicate ($X=a.sil.$) from \cite{dra85}, and amorphous SiO$_2$ ($X=silica$) from \cite{phi85} together with the system spectral response curve of the {\it{AKARI}}/IRC NIR/N3, MIR-S/S7, S9W, and S11 bands. While $Q^{abs}_{a.car.}(\lambda)$ shows a smooth fetureless profile without any band structures, $Q^{abs}_{a.sil.}(\lambda)$ and $Q^{abs}_{silica}(\lambda)$ show the band structure peaking at $\sim$9.5--10.0$\mu$m and $\sim$8.5--9.0$\mu$m, respectively. Therefore, these band structures intrinsic to the silicate-related dust could contribute to the excess emission in S7, S9W, and S11 bands.\\

\begin{figure}[htbp!]
\hspace{1.5cm}
\includegraphics[width=0.75\linewidth]{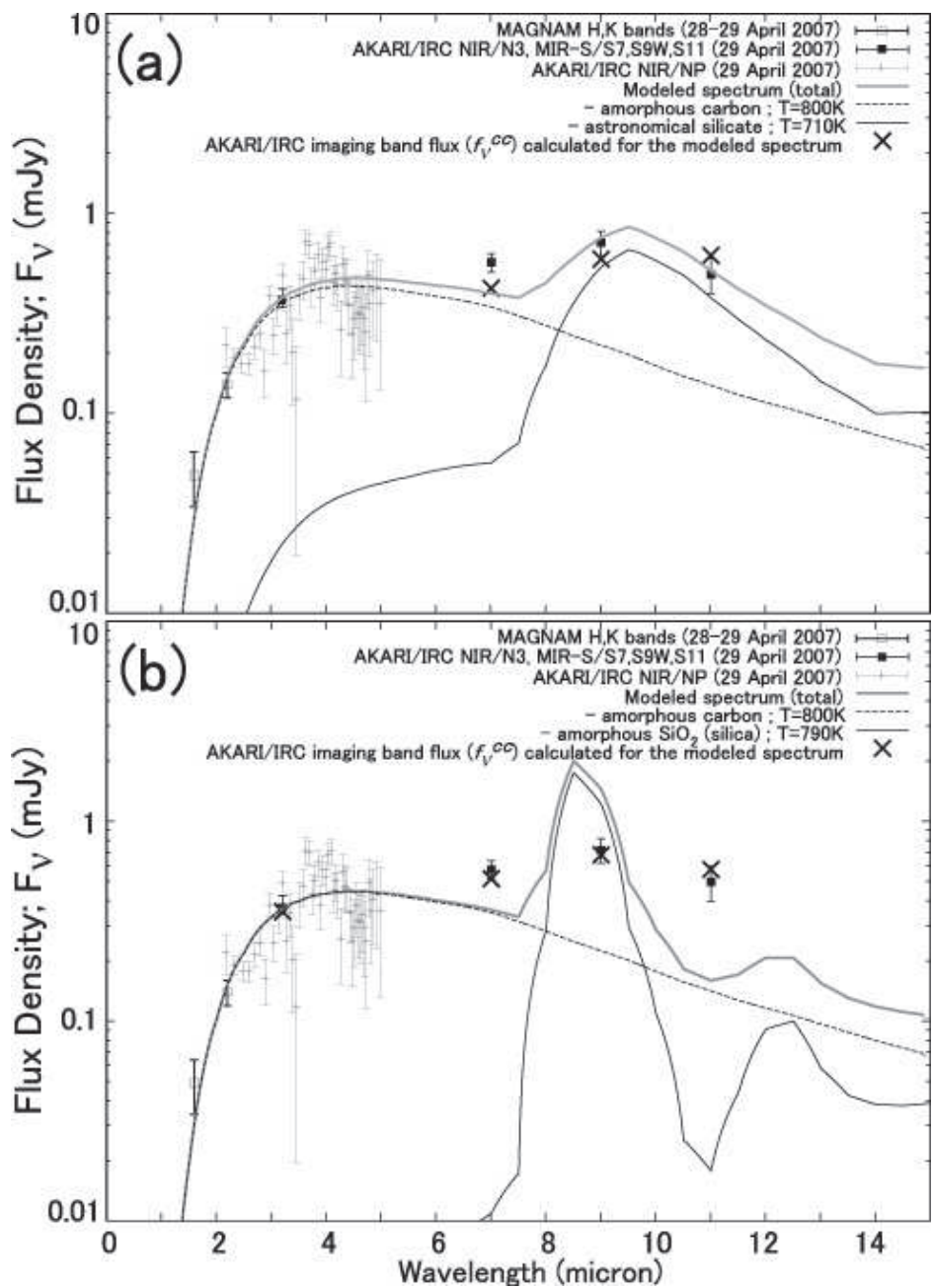}
\caption{The best fit model spectra with Eq.~(\ref{EQ3}) for the case of two component ($N=2$) models made of (a) amorphous carbon ($X_1=a.car.$) and astronomical silicate ($X_2=a.sil.$) and (b) amorphous carbon ($X_1=a.car.$) and amorphous SiO$_{2}$ ($X_2=silica$). Symbols are the same as in Fig.~\ref{f3}. \label{f5}}
\end{figure}

Firstly, we assume a two component ($N=2$) model of the amorphous carbon ($X_1=a.car.$) and the astronomical silicate ($X_2=a.sil.$) and the near- to mid-infrared data of SN2006jc on day 200 are fitted with Eq.~(\ref{EQ3}), where $T_{a.car.}=800$K and $M_{a.car.}=6.9\times 10^{-5}M_{\odot}$ are fixed and $T_{a.sil.}$ and $M_{a.sil.}$ are taken as the free parameters. The grain radii of both components are set as 0.01$\mu$m. The best fit model spectrum is shown in Fig.~\ref{f5}a and we obtain $T_{a.sil.}=710\pm 10$K and $M_{a.sil.}=1.12 \pm 0.02 \times 10^{-4}M_{\odot}$. \\

 However, the flux densities in S7, S9W and S11 bands predicted for this two component (amorphous carbon + astronomical silicate) model taking account of the color correction cannot well reproduce the observed mid-infrared SED characterised by the enhanced S9W flux density (see Fig.~\ref{f5}a). For one thing, the band structure of astronomical silicate peaking at $\sim$9.5--10.0$\mu$m is almost fully included not only in S9W but also, sometimes more efficiently depending on the dust temperature, in S11. For another, the predicted model spectrum is too low in $6<\lambda<8.5$$\mu$m, where the S11 band does not have its sensitivity but the S9W band does. Therefore, we need to have another dust component that has a band structure in $\lambda<8.5$$\mu$m in its absorption efficiency profile. Amorphous SiO$_2$ (silica) is one of the candidates for such component (see Fig.~\ref{f4}c). Moreover, silica should be able to be produced in SN ejecta \citep{noz03,noz07}.  Its presence in the Cassiopeia A supernova remnant has been reported by a mid-infrared spectroscopic observation with Short Wavelength Spectrometer (SWS) on board the Infrared Space Observatory (ISO) \citep{dou01} and also by a recent {\it{Spitzer}} observation \citep{rho07}.\\
 
 We assume a two component ($N=2$) model consisting of hot amorphous carbon ($X_1=a.car.$) and amorphous silica ($X_2=silica$), where $T_{a.car.}=800$K and $M_{a.car.}=6.9\times 10^{-5}M_{\odot}$ are fixed and $T_{silica}$ and $M_{silica}$ are taken as the free parameters. The grain radii of both components are set to be $0.01\mu$m and we assume the density of amorphous SiO$_2$ $\rho_{silica}=2.62$g cm$^{-3}$. The best fit model spectrum is shown in Fig.~\ref{f5}b and we obtain $T_{silica}=790\pm 10$K and $M_{silica}=6.5 \pm 0.1 \times 10^{-5}M_{\odot}$. This time, the predicted flux densities are in good agreement with the mid-infrared SED characterized by the enhanced S9W flux density. \\

 \pagebreak

\subsection{Two Temperature Amorphous Carbon Dust Model}
\begin{figure}[htbp!]
\hspace{1cm}
\includegraphics[width=0.85\linewidth]{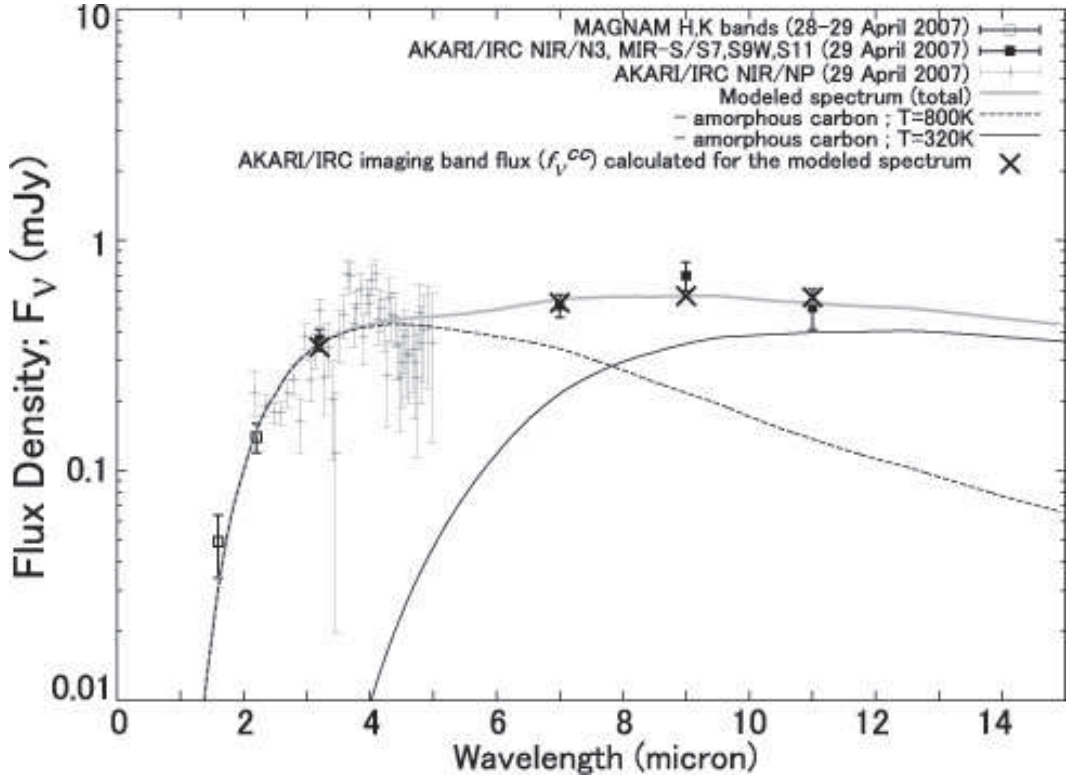}
\caption{The results of the best fit spectrum with Eq.~(\ref{EQ3}) are shown for the case of two component ($N=2$) models made of two temperature amorphous carbons ($X_1=h.a.car.$ and $X_2=w.a.car.$). Symbols are the same as in Fig.~\ref{f3}.
 \label{f6}}
\end{figure}

Another possible candidate for the excess component is the thermal emission from the featureless dust with temperature lower than 800K. Our near- to mid-infrared data of SN2006jc on day 200 are fitted with Eq.~(\ref{EQ3}) assuming a two component ($N=2$) model of hot amorphous carbon ($X_1=h.a.car.$) and warm amorphous carbon ($X_2=w.a.car.$). The temperature and the integrated mass of the former component are fixed to the values obtained in \S 3.2 and those of the latter component, $T_{w.a.car.}$ and $M_{w.a.car.}$, are taken as the free parameters. The same properties of the absorption efficiency profile, the grain radius and the density are assumed both for the warm and hot components. The fitting is made so that the flux densities of the model spectrum at S7, S9W, and S11 bands that were converted into the AKARI/IRC calibration system should best reproduce the observed flux density at S7, S9W, and S11 bands. The best fit model spectrum is shown in Fig.~\ref{f6} and we obtain $T_{w.a.car.}=320\pm 10$K and $M_{w.a.car.}=2.7^{+0.7}_{-0.5}\times 10^{-3}M_{\odot}$.

We note that the observed near- to mid-infrared SED has double peaked characteristics with one peak being located in the near-infrared and the other peak in the mid-infrared, and that the model fit with two-temperature amorphous carbon dust results in a considerable temperature gap between the two. This temperature gap can be explained if we assume the situation such that only the hot component with the temperature of 800K is the newly formed dust in the ejecta of SN2006jc and that the 320K warm component corresponds to thermal emission from the pre-existing circumstellar dust farther away from the newly formed dust with a dust-depleted region in between. The dust-depleted region may have been created by the dust evaporation due to the shock breakout in the initial phase of the SN explosion.

\pagebreak

\section{Discussion}
Observational evidence for the presense of silicate and/or slica grains formed in SN ejecta has been reported for the Cas A supernova remnant (\citealt{dou01,rho07}, and references therein), and the presense of silicate dust in the circumstellar medium is confirmed for SN 1987A \citep{bou06}. The progenitor of type-Ib SN 2006jc is considered to be a Wolf-Rayet star, suffering the mass loss during its late evolution and forming a dense circumstellar medium rich in He \citep{fol07}. The He layer of core collapse supernovae is suggested to contain more carbon than oxygen \citep{nom88} and, especially, the Wolf-Rayet star with strong mass loss leads to thick C-rich CSM and envelope \citep{lim06,tom07}. Therefore, the formation of silicate and/or silica grains in the C-rich circumstellar medium of SN2006jc is less likely (e.g., \citealt{noz03}).

Even if the silicate and/or silica dust is newly formed in the ejecta of SN2006jc, their temperature reaches only up to 100K taking account of the optical absorption and the collision with ejecta gas with 1000K on day $\sim$200 after the discovery \citep{noz07}. {\it{AKARI}}/IRC S7/S11 band colors calculated for astronomical silicate and silica with 100K are $f^{a.sil}_{\nu}(S7)/f^{a.sil.}_{\nu}(S11)=5.2\times 10^{-3}$ and $f^{silica}_{\nu}(S7)/f^{silica}_{\nu}(S11)=4.1\times 10^{-2}$, respectively, which are by far smaller than the S7/S11 band color of $\sim 0.5\pm 0.2$ obtained for the observed mid-infrared excess component. Therefore, both the astronomical silicate and amorphous silica with 100K cannot reproduce the flux density at S7, S9W, and S11 bands of the excess component and should not be the major carriers of the excess. Consequently, the mid-infrared excess component over the model spectrum of the amorphous carbon of 800K is likely to be IR light echo by amorphous carbon with 320$\pm$10 K in the circumstellar medium discussed in \S 4.2. However, the MIR photometric data with S7, S9W, and S11 bands are insensitive to the newly formed silicate of less than $\sim 1 M_{\odot}$ as long as the tempearute is $\sim 100$ K. In this sense, the 6.9$\times 10^{-5}M_{\odot}$ of amorphous carbon should be a lower limit of the newly formed dust mass in the ejecta of SN 2006jc.

The formation scenario of silicate or silica dust in SN ejecta is illustrated as the followings; As the oxygen-rich ejecta expands and the gas temperature becomes lower than $\sim 3000$K, silicon monoxide molecules are formed. Then, when the SiO molecular gas is cooled to 1200--1500 K, chemical reactions to condense silicate or silica dust become active \citep{noz03}. Observational evidence for SiO molecules being the precursors of silicate or silica dust in supernova SN1987A is reported by \cite{roc91}, in which they find the coincidence of the decrease in the mass of SiO molecules with the onset of the dust emission from the supernova. \cite{kot06} also report the presence of SiO molecules in a Type IIP supernova 2005af based on mid-infrared observations with the $Spitzer$ $Space$ $Telescope$. 

We still find a small amount of excess in the S9W band over the best fit spectrum for the two temperature amorphous carbon dust model (see Fig.~\ref{f6}). SiO molecules carry the $\Delta v=1$ vibrational-rotational bands extending from 7.5--9.5$\mu$m \citep{bee74}, which are more efficiently included in the S9W band than in the S7 and S11 bands. If the oxygen-rich part resides in the SN ejecta gas whose temperature is still higher than the condensation temperature of silicate and silica dust ($\sim$1200--1500 K), precursory SiO molecules can account for this excess in the S9W band. 

Finally, we conclude that amorphous carbon of $800\pm 10$K of $6.9\pm 0.4 \times 10^{-5} M_{\odot}$ is newly formed in the ejecta of SN2006jc on day 200. This dust mass is more than 3 orders of magnitude smaller than the amount needed for core collapse supernovae to contribute efficiently to the early-Universe dust budget \citep{mor03}. Recent observational studies report the evidence of dust formation in the ejecta of several core collapse supernovae, but the produced dust mass in the SN ejecta is generally found to be much smaller than theoretically predicted values of 0.1-1$M_{\odot}$ \citep{noz03}, i.e. no more than 7.5$\times 10^{-4}M_{\odot}$ for SN 1987A \citep{erc05}, $\sim 10^{-4}M_{\odot}$ for SN 1999em \citep{elm03}, and 4$\times 10^{-5}M_{\odot}$ for SN 2003gd \citep{mei07}. The obtained amorphous carbon dust mass for SN2006jc on day 200 in our study is consistent with those relatively small values obtained for other several dust forming core collapse supernovae, although our data is almost insensitive to the newly formed silicate of $\sim$100 K. It should also be noted that the dust mass is usually derived by assuming that dust emission is optically thin, except for \cite[][]{erc05}, and the derived mass is considered to be a lower limit of the dust mass \citep[][]{mei07}. In addition to the dust formed in the ejecta, on the other hand, our mid-infrared photometric data suggest the presence of another amorphous carbon dust of $320\pm 10$K of $2.7^{+0.7}_{-0.5} \times 10^{-3} M_{\odot}$ as the circumstellar component. This component is expected to be formed in the mass loss wind associated with the Wolf-Rayet stellar activity (e.g., \citealt{wil92,wat97,mol99,voo00}). It follows that the dust condensation not only in the SN ejecta itself but also in the mass loss wind associated with the prior events to the SN explosion could make a significant contribution to the dust formation by a massive star in its whole evolutional history.

\section{Summary}
 We present our latest results on near- to mid- infrared observations of supernova (SN) 2006jc at 200 days after the discovery using the Infrared Camera (IRC) on board $AKARI$. The near-infrared (2--5$\mu$m) spectrum of SN 2006jc is obtained for the first time and is found to be well interpreted in terms of the thermal emission from amorphous carbon of 800$\pm 10$K of $6.9\pm 0.5 \times 10^{-5}M_{\odot}$ that was formed in the SN ejecta. This newly formed dust mass in the SN ejecta is in a range similar to those obtained for other several dust forming core collapse SNe based on recent observations. Mid-infrared photometric data with {\it{AKARI}}/IRC MIR-S/S7, S9W, and S11 bands have shown excess emission over the thermal emission of amorphous carbon of 800K. This mid-infrared excess emission is likely to be accounted for by the emission from circumstellar amorphous carbon dust of 320$\pm 10$K of 2.7$^{+0.7}_{-0.5} \times 10^{-3}M_{\odot}$ rather than by the band emission of astronomical silicate and/or silica grains. Since this circumstellar amorphous carbon dust of 320K is expected to be formed in the mass loss wind associated with the Wolf-Rayet stellar activity, our result suggests that a significant amount of dust is condensed in the mass loss wind associated with the prior events to the SN explosion. Finally, a possible contribution of emission bands by the precursory SiO molecules in 7.5--9.5$\mu$m is suggested to explain the enhanced S9W flux density. Mid-infrared spectroscopy of early-time SNe of similar type is highly important for further and detailed analysis on the properties of newly formed dust in the supernova ejecta distinguishing from the circumstellar or interstellar dust component.

This work is based on observation of {\it{AKARI}}, a JAXA project with the participation of ESA. We thank all the members of the {\it{AKARI}} project, particularly those who have engaged in the observation planning and the satellite operation during the performance verification phase, for their continuous help and support. We would also express our gratitude to the {\it{AKARI}} data reduction team for their extensive work in developing data analysis pipelines. This work is supported in part by a Grant-in-Aid for Scientific Research on Priority Areas from the Ministry of Education, Culture, Sports, Science, and Technology of Japan and Grants-in-Aid for Scientific Research from the JSPS.


\begin{thebibliography}{}
\bibitem[Arkharov et al.(2006)]{ark06}Arkharov,A., et al.2006, ATel, 961, 1
\bibitem[Beer, Lambert \& Sneden(1974)]{bee74}Beer, R., Lambert, D. L., Sneden, C., 1974, PASP, 86, 806
\bibitem[Bouchet et al.(2006)]{bou06}Bouchet, P., Dwek, E., Danziger, J., Arendt, R. G., De Buizer, I. J. M., Park, S., Suntzeff, N. B., Kirshner, R. P., Challis, P., 2006, \apj, 650, 212
\bibitem[Draine(1985)]{dra85} Draine,B.T. 1985, \apjs, 57, 587
\bibitem[Douvion, Lagage \& Pantin(2001)]{dou01}Douvion,T., Lagage,P.O., \& Pantin,E., 2001, \aap, 369, 589
\bibitem[Edo(1983)]{edo83} Edo,O. 1983, PhD thesis, Univ. of Arisona
\bibitem[Elmhamdi et al.(2003)]{elm03}  Elmhamdi, A., Danziger, I. J., Chugai, N., Pastorello, A., Turatto, M., Cappellaro, E., Altavilla, G., Benetti, S., Patat, F., Salvo, M., 2003, \mnras, 338, 939
\bibitem[Elmhamdi et al.(2004)]{elm04}Elmhamdi, A., Danziger, I. J., Cappellaro, E., Della Valle, M., Gouiffes, C., Phillips, M. M., Turatto, M., 2004, \aap, 426, 963
\bibitem[Ercolano, Barlow \& Storey(2005)]{erc05} Ercolano,B., Barlow,M.J., Storey,P.J., 2005, MNRAS, 362, 1038
\bibitem[Foley et al.(2007)]{fol07}Foley,R.J., Smith,N., Ganeshalingam,M., Li,W., Chornock,R., and Filippenko,A.V., 2007, \apj, 657, L105
\bibitem[Hines et al.(2004)]{hin04} Hines, D. C., et al. 2004, \apjs, 154, 290
\bibitem[Immler, Modjaz \& Brown(2006)]{imm06}Immler,S., Modjaz,M., \& Brown,P.J. 2006, ATel, 934, 1
\bibitem[Kobayashi et al.(1998a)]{kob98a}Kobayashi,Y., et al.1998, Proc. SPIE, 3352, 120
\bibitem[Kobayashi et al.(1998b)]{kob98b}Kobayashi,Y., et al.1998, Proc. SPIE, 3354, 769
\bibitem[Kotak et al.(2006)]{kot06}Kotak,R., et al. 2006, \apj, 651, 117
\bibitem[Kozasa, Hasegawa \& Nomoto(1991)]{koz91} Kozasa,T., Hasegawa,H., Nomoto,K., 1991, \aap, 249, 474
\bibitem[Limongi \& Chieffi(2006)]{lim06} Limongi, M. \& Chieffi, A., 2006, \apj, 647, L483
\bibitem[Meikle et al.(2007)]{mei07} Meikle,W.P.S., et al.2007, \apj, 665, 608
\bibitem[Minezaki, Yoshii, \& Nomoto]{min07}Minezaki,T., Yoshii,Y., \& Nomoto,K. 2007, IAUC, 8883
\bibitem[Molster et al.(1999)]{mol99} Molster,F.J., et al. 1999, \aap, 350, 163
\bibitem[Morgan \& Edmunds(2003)]{mor03}Morgan,H.L., Edmunds,M.G., 2003, MNRAS, 343, 427
\bibitem[Murakami et al.(2007)]{mur07} Murakami,H., et al.2007, \pasj, 59, S369
\bibitem[Nakano et al.(2006)]{nak06} Nakano,S., Itagaki,K., Puckett,T., \& Gorelli,R. 2006, Cent. Bur. Electron. Tel., 666, 1
\bibitem[Nomoto \& Hashimoto(1988)]{nom88}Nomoto,K., \& Hashimoto,M., Phys.Rep.163, 13
\bibitem[Nozawa et al.(2003)]{noz03} Nozawa,T., Kozasa,T., Umeda,H.,Maeda,K., \& Nomoto,K. 2003, \apj, 598, 785
\bibitem[Nozawa et al.(2007)]{noz07} Nozawa,T., et al.2007, in preparation
\bibitem[Ohyama et al.(2007)]{ohy07}Ohyama,Y., et al.2007, \pasj, 59, S411
\bibitem[Onaka et al.(2007)]{ona07} Onaka,T., et al.2007, \pasj, 59, S401
\bibitem[Pastorello et al.(2007)]{pas07} Pastorello, A., et al.2007, nature, 447, 829 
\bibitem[Philipp(1985)]{phi85} Philipp,H.R. 1985, in Palik,E.D. ed., Handbook of Optical Constants of Solids. Academic Press, San Diego, P719
\bibitem[Roche et al.(1991)]{roc91}Roche, P. F., Aitken, D. K., Smith, C. H., 1991, \mnras, 252, 39
\bibitem[Rho et al.(2007)]{rho07}Rho,J., Kozasa,T., Reach, W.T., Rudnick,L., DeLaney,T., Smith,J.D., Ennis,J.A., Gomez,H., Tappe,A., 2007, \apj, in press [arXiv: astro-ph/07092880]
\bibitem[Sakon et al.(2007)]{sak07} Sakon,I., et al.2007, \pasj, 59, S483
\bibitem[Smith et al.(2007)]{smi07} Smith,N., Foley,R.J., \& Filippenko,A.V. 2007, \apj, submitted [arXiv: astro-ph/07042249]
\bibitem[Sugerman et al.(2006)]{sug06} Sugerman,B.E.K., et al.2006
\bibitem[Todini \& Ferrara(2001)]{tod01} Todini,P., Ferrara,A., 2001, MNRAS, 325, 726
\bibitem[Tominaga et al.(2007)]{tom07}Tominaga, N., et al. 2007, \apj, submitted [arXiv: astro-ph/0711.4782]
\bibitem[Voors et al.(2000)]{voo00} Voors,R.H.M., et al. 2000, \aap, 356, 501
\bibitem[Waters et al.(1997)]{wat97} Waters,L.B.F.M., Morris,P.W., Voors,R.H.M., Lamers,H.J.G.L.M., Trams,N.R. 1997, Ap\&SS, 255, 179
\bibitem[Williams et al.(1992)]{wil92} Williams, P.M., et al. 1992, \mnras, 258, 461
\end{thebibliography}
\end{document}